\documentclass[conference]{IEEEtran}
\IEEEoverridecommandlockouts
\usepackage{cite}
\usepackage{amsmath,amssymb,amsfonts}
\usepackage{algorithmic}
\usepackage{graphicx}
\usepackage{textcomp}
\usepackage[linesnumbered,ruled,vlined]{algorithm2e}

\usepackage[letterpaper, top=.75in, bottom=1in, left=0.625in, right=0.625in]{geometry}
\usepackage{xcolor}
\def\BibTeX{{\rm B\kern-.05em{\sc i\kern-.025em b}\kern-.08em
    T\kern-.1667em\lower.7ex\hbox{E}\kern-.125emX}}
\begin{document}

\title{Smart Pilot Assignment for IoT in Massive MIMO Systems: A Path Towards Scalable IoT Infrastructure
\\
\thanks{This research was partly funded by Palmer Department Chair Endowment Funds at Iowa State University.}
}

\author{\IEEEauthorblockN{ Muhammad Kamran Saeed}
\IEEEauthorblockA{\textit{Department of Electrical and}\\\textit{ Computer Engineering} \\
\textit{Iowa State University}\\
Ames, USA \\
Kamran@iastate.edu}
\and

\IEEEauthorblockN{Ashfaq Khokhar}
\IEEEauthorblockA{\textit{Department of Electrical and}\\\textit{ Computer Engineering} \\
\textit{Iowa State University}\\
Ames, USA \\
Ashfaq@iastate.edu}

}

\maketitle

\begin{abstract}

5G sets the foundation for an era of creativity with its faster speeds, increased data throughput, reduced latency, and enhanced IoT connectivity, all enabled by Massive MIMO (M-MIMO) technology. M-MIMO boosts network efficiency and enhances user experience by employing intelligent user scheduling. This paper presents a user scheduling scheme and pilot assignment strategy designed for IoT devices, emphasizing mitigating pilot contamination, a key obstacle to improving spectral efficiency (SE) and system scalability in M-MIMO networks. We utilize a user clustering-based pilot allocation scheme to boost IoT device scalability in M-MIMO systems. Additionally, our smart pilot allocation minimizes interference and enhances SE by treating pilot assignment as a graph coloring problem, optimizing it through integer linear programming (ILP). Recognizing the computational complexity of ILP, we introduced a binary search-based heuristic predicated on interference threshold to expedite the computation, while maintaining a near-optimal solution. The simulation results show a significant decrease in the required pilot overhead (about 17\%), and substantial enhancement in SE (about 8-14\%).

\end{abstract}


\section{Introduction}

5G paves the way for a new era of innovation, offering higher throughput, lower latency, and enhanced IoT connectivity. M-MIMO is the backbone of 5G, enhancing faster data speeds, enabling multi-device connectivity, improving network efficiency, and more. Obtaining precise channel state information (CSI) at the base station is a pre-requisite for harnessing the benefits of M-MIMO systems\cite{intro2}. The standard practice to acquire CSI is through a pre-defined training sequence called pilot signal. Typically, orthogonal pilot signals are allocated to individual users within each cell to enable the differentiation of the received signal of distinct users at the base station. However, using orthogonal pilot sequences is constrained by the limited coherence block length, ultimately impacting the system's scalability. This leads to the reuse of pilot sequences in neighboring cells, making it challenging for the base station to differentiate between pilot signals from neighboring cells and within a cell. This phenomenon, known as inter-cell pilot contamination, serves as a significant hurdle in the quest to enhance the spectral efficiency  (SE) of M-MIMO systems\cite{intro3}. 

\par To overcome the mentioned challenges, M-MIMO employs intelligent user scheduling techniques to dynamically assign resources, aiming to maximize SE while enhancing user gratification. User scheduling involves the carefully chosen subsets of users to reduce mutual interference during data transmission. In \cite{intro1}, the author introduces a user scheduling scheme to organize users while minimizing interference, enabling resource-efficient sharing without imposing group size restrictions. Thus, high device volumes in groups may introduce potential delays as devices may need to wait their turn to transmit data. This raises questions about the system's capacity to ensure effective data transmission while adhering to timing requirements.

\par To overcome this challenge, our prior research\cite{my} introduced a user scheduling strategy designed to mitigate inter-cell pilot contamination by employing a max K-cut partitioning method. The proposed scheme effectively integrated several devices; however, it also encountered several limitations. Firstly, the max K-cut problem is NP-hard, which makes finding optimal solutions computationally demanding. Secondly, devices unable to transmit in specified intervals due to poor channel conditions are termed as 'omitted devices'. These devices will likely not have the chance to transmit. Lastly, our prior proposed method assumed device homogeneity, with identical data sizes and transmission intervals for all devices. These assumptions limit the applicability of the solution proposed in~\cite{my} to just a few real-world scenarios.

\par In this paper, we address the above-described issues in the prior state of the art. Firstly, we propose an integer linear programming (ILP)-based pilot assignment scheme to mitigate pilot contamination in IoT-based M-MIMO systems. ILP delivers precise, discrete solutions in optimization problems, though a notable constraint lies in the time it consumes. To tackle this challenge, we propose a binary search-based heuristic to find a trade-off between attaining a solution that approaches optimality while minimizing the computation time. Secondly, we considered heterogeneous devices to better reflect real-world scenarios. Lastly, in the proposed solution, every device gets an opportunity to transmit, regardless of its channel conditions. Simulation results illustrate enhanced SE and scalability for IoT devices in M-MIMO systems.  

\section{Contribution}
The key contributions can be outlined as follows:
\begin{enumerate}

    \item In addressing scalability concerns, we introduce a user grouping-based pilot assignment scheme that enables the seamless integration of a growing number of IoT devices,  while optimizing the resource-efficient utilization of scarce pilot signals in the M-MIMO system. 
    \item We reformulated the pilot contamination as a problem of graph coloring and optimized its solution using ILP.
    \item To tackle the time-consuming nature of the ILP method, we propose a binary search-based heuristic predicated on interference threshold to strike a balance between computation time and achieving solutions close to being optimal.

   \item We have incorporated heterogeneous devices to closely align with real-world situations without any device being omitted due to adverse channel conditions.

\end{enumerate}

\section{System Model}

Inspired by our prior research \cite{my}, this system model showcases familiar elements. For the sake of completeness, we briefly discuss the CSI acquisition and basic signal processing. This study evaluates an uplink M-MIMO network spanning L cells. Each cell contains \( K \) IoT devices, grouped into \( C \) clusters by spatial correlation with \( U_d \) devices per cluster, and each cell's base station has M antennas. The communication channel between the \( k_{\text{th}} \) device of \( i_{\text{th}} \) cell (with \( i \) ranging from 1 to \( L \)) and its base station is \( g^i_{ik} \in \mathbb{C}^M \). In contrast, the interference channel between the \( k_{\text{th}} \) device in the \( j_{\text{th}} \) cell and the \( i_{\text{th}} \) cell's base station is \( g^i_{jk} \in \mathbb{C}^M \), as shown in Fig. 1. 

\vspace{-0.3cm}
\begin{figure}[htp]
    \centering
    \includegraphics[width=8.65cm]{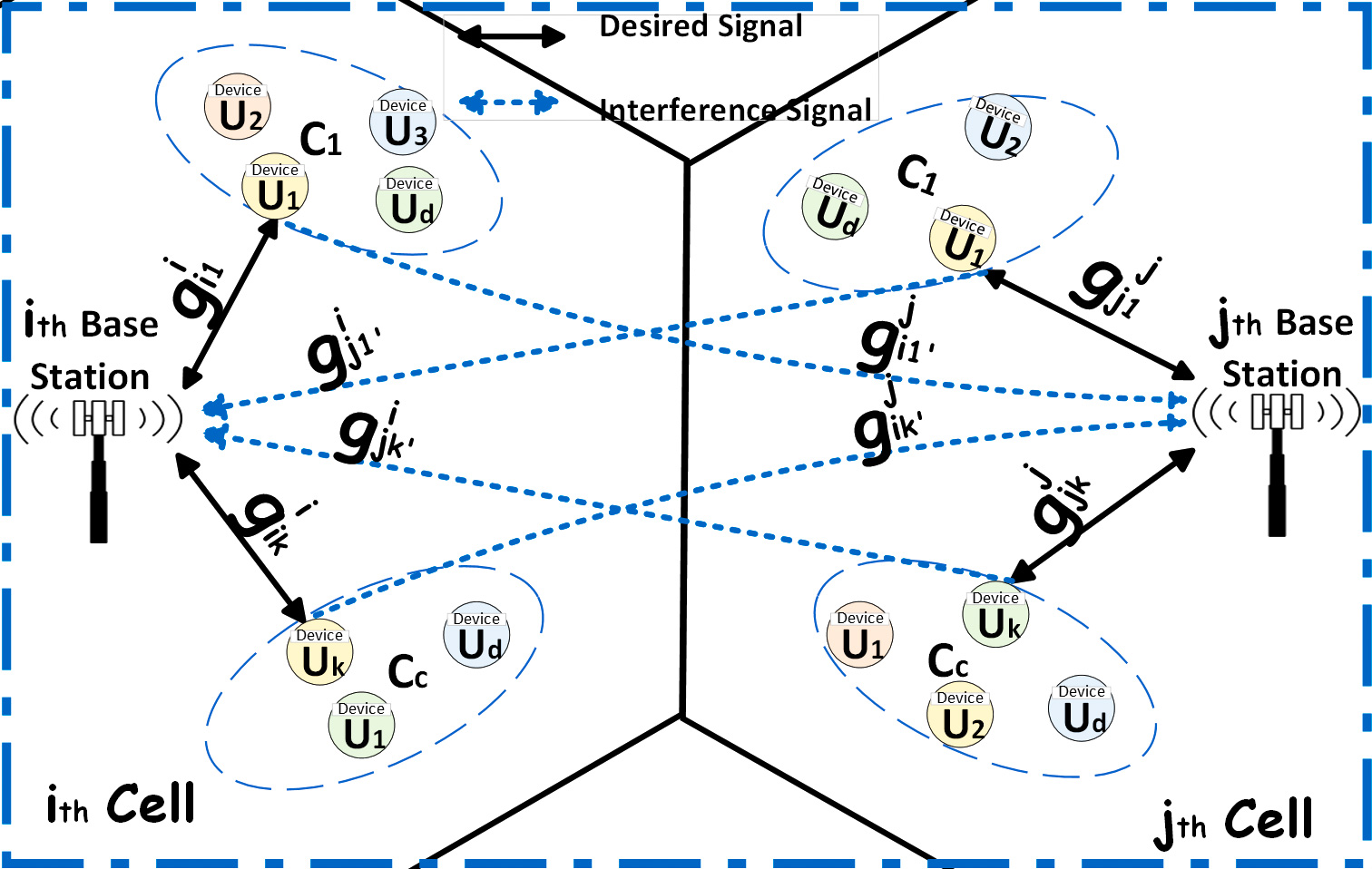}
    \vspace{-0.2cm}
    \caption{Desired and interfering signals from cells $i_{\text{th}}$ and $j_{\text{th}}$
}
    \label{fig:pilot}
\end{figure}

\par In this paper, we consider spatially correlated Rayleigh fading channels. A convenient way to represent spatially correlated Rayleigh fading channels in the absence of a line of sight can be expressed as $\mathbf {g}_{ik}^{j} \sim \mathcal {N}_{\mathbb {C}} \left ({\mathbf {0}_{M}, \mathbf {R}_{ik}^{j} }\right)$. The average channel gain $\mathbf {\beta}_{ik}^{j}$ of an antenna at $j_{th}$ base station for $k_{th}$ device of $i_{th}$ cell is given by $\frac{1}{M}tr(\mathbf {R}_{ik}^{j})$ \cite{intro2}.

\subsection{Channel State Information}

\par
\begin{figure}[htp]
    \centering
    \includegraphics[width=8.65cm]{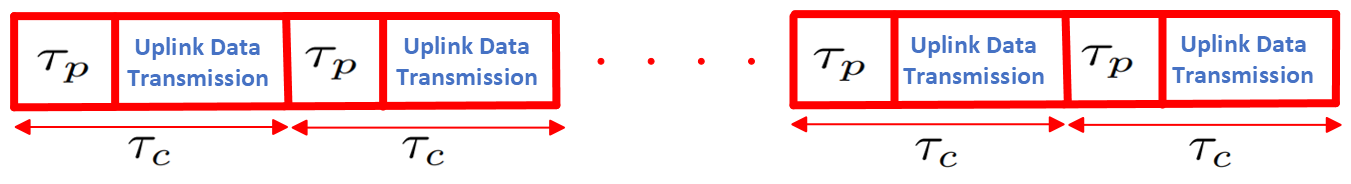}
   \vspace{-0.3cm}
    \caption{Coherence Intervals}
    \label{fig:coherence}
\end{figure}

 Time-frequency are segmented into smaller blocks (coherence intervals), represented as \( \tau_{c} \), where channel responses are roughly constant. A segment \( \tau_{\rho} \) is allocated within \( \tau_{c} \) for uplink training with pre-defined pilot symbols as shown in Fig. \ref{fig:coherence}.

\vspace{-0.3cm}
\vspace{-0.3cm}

Pilot symbols enable the base station to obtain the  \( k_{th} \) device's channel state information, denoted as \( {\hat {g}_{ik}^j} \). The pilot signals received at the \( i_{th} \) base station are expressed as:

\begin{equation}
{\mathbf Y{{'}_{i}}=   {\sum _{i=1}^{L} \sum _{k=1}^{K} \sqrt { p_{ik}}  \mathbf g_{ik}^{i} \boldsymbol{\Phi }_{ik}^{H}} + {\mathbf N_{{i}}}}
\end{equation}

The pilot sequence for the \( k_{\text{th}} \) device in the \( i_{\text{th}} \) cell is \( \boldsymbol{\Phi}_{ik}^{H} \) and is transmitted with power \( p_{ik} \). The noise, \( N_i \), represents additive receiver noise following a complex Gaussian distribution \( \mathcal{N}_\mathbb{C} \left(0, \sigma_{\text{ul}}^2 I_M\right) \). The received pilot signal is despread by multiplying with the predefined pilot sequence of the \( k^{\text{th}} \) device.

\begin{align*}
\mathbf y_{ik} &= \mathbf{Y{'}}_{i} \boldsymbol{\Phi}_{ik} = {\sum_{k=1}^{K}\sqrt{p_{ik}} \mathbf{g}_{ik}^{i} \boldsymbol{\Phi}_{ik}^{H}}\boldsymbol{\Phi}_{ik} \\
&\quad+ {\sum_{j=1,j\neq i}^{L} \sum_{k=1}^{K} \sqrt{p_{jk}} \mathbf{g}_{jk}^{i} \boldsymbol{\Phi}_{jk}^{H}}\boldsymbol{\Phi}_{ik} + {\mathbf{N}_{i}}\boldsymbol{\Phi}_{ik} \tag {2}
\end{align*}

Typically, orthogonal pilot sequences are assigned to devices within a cell and reused in other cells. Hence, \( \boldsymbol{\Phi }_{ik} \boldsymbol{\Phi }_{jk'}^{H} = \tau_{\rho} \), if devices \( k \) and \( k' \) with identical pilot sequences in cells \( i \) and \( j \); otherwise, it is 0.

 \begin{align*}
\mathbf y_{ik} = \underbrace{\sum_{k=1}^{K}\sqrt{p_{ik}} \mathbf{g}_{ik}^{i} \tau_{\rho}}_{\mathrm{Desired~channel}} 
&+ \underbrace{\sum_{j=1,j\neq i}^{L} \sum_{k=1}^{K} \sqrt{p_{jk}} \mathbf{g}_{jk}^{i} \tau_{\rho}}_{\mathrm{Interfering~ channel}} + \underbrace{\mathbf{N}_{i} {\Phi}_{ik}}_{\mathrm{Noise}} \tag {3}
\end{align*}

 The 'desired channel' is the channel for the \(k_{th}\) user in the \(i_{th}\) cell received by its base station. In contrast, the second term represents interference from \(k_{th}\) devices in the \(j_{th}\) cell employing the same pilot as the \(i_{th}\) cell. We estimate the channel for \({g}_{ik}^{i}\) with an MMSE estimator, resulting in the vector \(\hat { {g}}_{ik}^{i}\) that minimizes the squared error.

\begin{align} \hat {  {g}}_{ik}^{i} =& \mathbf{R}_{ik}^{i} \big ( \mathbf {Q}_{ik}^{i}  \big)^{-1} \mathbf{y}_{ik}=& \mathbf{R}_{ik}^{i} \big ( {\sum _{j=1}^{L} } \mathbf {R}_{jk}^{i}+\frac {1}{\tau _{p}}\frac {\sigma _{\mathrm{ul}}^{2}}{{\sqrt {p_{ik}}}} \mathbf {I}_{M}  \big)^{-1} \mathbf{y}_{ik} \tag {4}
\end{align}

\( Q_{ik}^{i} \) is the inverse of the normalized correlation matrix for the processed received signal. The channel estimation error, \( \Tilde{ g}_{ik}^{i} = { g}_{ik}^{i} - \hat { {g}}_{ik}^{i} \), is given by:

\begin{align} \frac {\mathbb {E}\{ \| \mathbf {g}_{ik}^{i} - \hat { \mathbf {g}}_{ik}^{i} \|^{2} \}}{\mathbb {E}\{ \| \mathbf {g}_{ik}^{i} \|^{2} \}}=\frac {{ \mathrm {tr}{\big (\mathbf {R}_{ik}^{i}\big)}} - \mathrm {tr}{\big (\mathbf {R}_{ik}^{i} \big (\mathbf {Q}_{ik}^{i}\big)^{-1} \mathbf {R}_{ik}^{i}\big)}}{ \mathrm {tr}{\big (\mathbf {R}_{ik}^{i}\big)}} \tag {5}
\end{align}

The normalized minimum squared estimation error is affected by pilot contamination (PC). Pilot signals from neighboring cells disrupt channel estimates in the \(i_{\text{th}}\) cell, posing a challenge in M-MIMO systems.

\subsection{Signal Processing at Base Station}
In the context of uplink data transmission, the signal received at the $i_{th}$ base station can be expressed as:    
\begin{equation} 
{\mathbf {Y}_{i} =  {  \sum _{k=1}^{K} \sqrt{p_{ik}} \mathbf {g}_{ik}^{i} \mathbf x_{ik}} +   { \sum _{j=1,j \neq i}^{L} \sum _{k=1}^{K} \sqrt{p_{jk}} \mathbf {g}_{jk}^{i} \mathbf{x}_{jk}} +  {\vphantom {\sum _{i=1,i\ne k}^{K} } \mathbf {N}_{i}}} \tag {6}
\end{equation}

\( \mathbf N_{i}\sim \mathcal {N}_{\mathbb {C}}(0, \sigma_{ul}^2 I_{M})\) is independent noise. \( \mathbf x_{ik} \) is the data signal from the \( k_{th} \) device in the \( i_{th} \) cell sent to its base station, with \( {p_{ik}} \) as the uplink power. The \( i_{th} \) cell's base station uses the combining vector \( \hat {V}^{i} = \mathbf{[\hat{v}_{i_{1}}, \hat{v}_{i_{2}},\dots, \hat{v}_{i_{k}} ]} \) to extract the target signal from interference and noise.

\begin{align*}
\hat {\mathbf v}^{H}_{ik} \mathbf Y_{i}=& {\sqrt{p_{ik}} \hat {\mathbf v}^{H}_{ik} \hat{\mathbf {g}}_{ik}^{i} \mathbf x_{ik}} + { \sum _{k'=1,k'\ne k}^{K} \sqrt{p_{ik'}} \hat{{\mathbf v}}^{H}_{ik} \hat{\mathbf {g}}_{ik'}^{i} \mathbf{x_{ik'}}} \\ &+  {{\sum _{\substack {j=1, j\neq i}}^{L}  \sum _{k=1}^{K}} \sqrt{p_{jk}} \hat{{\mathbf v}}^{H}_{ik}  \hat{\mathbf g}_{jk}^{i} \mathbf x_{jk}} +  {\hat{{\mathbf v}}^{H}_{ik}  \mathbf{N}_{ik}} \tag {7}
\end{align*}

 The uplink spectral efficiency $\mathsf {SE}_{ik}^{ul}$ can be calculated as follows:

\begin{equation} { \mathsf {SE}}_{ik}^{ul}= \big( \frac {\tau_{c} - \tau_{\rho}}{\tau _{c}} \big) \mathbb{E}
 \{\log _{2} (1 + {\gamma}_{ik})\}
 \tag {8} \end{equation}

\( \tau_c - \tau_\rho \) denotes the uplink data samples, while \( \gamma_{ik} \) represents the \( k_{\text{th}} \) device's SINR, defined as:

\begin{equation}
{{\gamma}_{ik} = \frac {\underbrace {{{p_{ik}}|\hat{{\mathbf v}}^{H}_{ik}} \hat{\mathbf g}_{ik}^{i}|^2}_{Desired}} {{p_{ik}}\hat{{\mathbf v}}^{H}_{ik} \left (  \underbrace{ \sum _{k'=1, k'\ne k}^{K} | \hat{{\mathbf g}}_{ik'}^{i} |^2   }_{intracell~ PC} +   \underbrace{{\sum _{\substack {j=1, j\neq i}}^{L} \sum _{k=1}^{K} |\hat{{\mathbf g}}_{jk}^{i}|^2 }}_{intercell ~PC} + Z_{i} \right )\hat{{\mathbf v}}^{H}_{ik}}} \tag {9}
\end{equation}

'Desired' is the signal from the \(k_{\text{th}}\) device. The term 'intracell PC' refers to variations within the cell, primarily influenced by channel estimation errors due to $k'$ devices. 'intercell PC' originates from the \(k_{\text{th}}\) device in the \(j_{\text{th}}\) cell employing the same pilot as the \(i_{\text{th}}\) cell. \( \mathbf {Z}_{i} \) is expressed as:

\begin{equation} \mathbf {Z}_{i} = \sum \limits _{j=1}^{L} \sum \limits _{k=1}^{K} (\mathbf {R}_{jk}^{i} - \mathbf {R}_{jk}^{i} \big (\mathbf {Q}_{jk}^{i}\big)^{-1} \mathbf {R}_{jk}^{i}) + \frac {\sigma _{\mathrm{ul}}^{2}}{\rho _{\mathrm{ul}}} \mathbf {I}_{M} \tag {10}   \end{equation}

The optimal receive-combining scheme reduces mean square error for \(s_{ik}\) and the received signal at the base station. Therefore, we explore the multicell M-MMSE combining scheme \({\mathbf {V}^{i}_{\mathrm{M-MMSE}}}\) for the \(i_{\text{th}}\) cell, defined as:

\begin{equation}
{\mathbf {V}^{i}_{\mathrm{M-MMSE}} = \left({\sum \limits _{j=1}^{L} \widehat { \mathbf {G}}_{j}^{i} {(\widehat { \mathbf {G}}_{j}^{i})}^{ {\mathrm {H}}} + \mathbf {Z}_{i} }\right)^{\!-1} \widehat { \mathbf {G}}_{i}^{i}} \tag {11}
\end{equation}

Where $ \hat{{G}}_{j}^{i} = \mathbf{[\hat{{g}}_{j1}^{i}, \hat{{g}}_{j2}^{i}, \ldots, \hat{{g}}_{jk}^{i} ]}$ represents the aggregation of individual channels of each device.

\section{Problem Formulation}

In this paper, we introduce an intelligent pilot assignment technique to enhance both the SE and system scalability of M-MIMO networks. Our approach incorporates a user grouping mechanism for IoT devices, aiming to reduce the pilot overhead. Moreover, we model the pilot assignment problem as a graph coloring problem and utilize ILP to allocate the pilot sequences to minimize inter-cell interference. Recognizing that ILP can be computationally intensive, we introduce an efficient heuristic based on the binary search-based method to find sub-optimal solutions within a given time frame.

\subsection{User Grouping}
Generally, every device uses its own orthogonal pilot sequence during uplink data transmission within a cell. In the uplink data transmission, a specific coherence block segment is allocated for pilot signal transmission, represented by $\tau _{\rho}$. Typically, we require a minimum of $K$ orthogonal pilot signals, where $\tau_{\rho} \ge K$, for each device to perform channel estimation. Thus, the demand for orthogonal pilot signals grows as the number of devices increases in a cell. Consequently, this increasing allocation of resources to pilot symbols transmission reduces the available space within coherence blocks for data transmission. This limitation poses a challenge to improving SE and the system's scalability within M-MIMO systems~\cite{my}.  

\par To address the challenge posed by the increasing demand for orthogonal pilot signals as the number of devices grows, in our previous work~\cite{my}, we introduced a clustering approach. In the proposed scheme, the authors made strategic use of a fundamental trait exhibited by the IoT devices, which is their inherent periodicity in transmitting data at predetermined and recurring intervals. This periodicity, contingent upon the data's criticality, can range from milliseconds to minutes. They employed a method of allocating distinct orthogonal pilots to each cluster to curtail the pilot signal requirements. Each device sequentially utilizes the same pilot sequence within each cluster during data transmission. The clustering of devices is achieved by utilizing the K-faster medoid clustering algorithm, with the additional complexity of $\mathbf{O(K(K - C))}$, which incorporates the spatial channel correlation of devices as a primary criterion for the clustering process. The spatial correlation similarity matrix can be computed as:

\begin{align} S_{(U_{k},U_{k'})}=\frac{\mathrm{Tr}\left({\mathbf R_{ik}^{i} \mathbf R^{iH}_{ik^{\prime }}}\right)}{\left|\left|\mathbf R_{ik}^{i}\right|\right|_F\left|\left|\mathbf R_{ik^{\prime }}^{i}\right|\right|_F},\;\; k\ne k^{\prime }, \tag {12} \end{align}

In the above equation, $U_k$ and $U_{k^{'}}$ are two distinct devices in the $i_{th}$ cell. However, the proposed work \cite{my} primarily investigates the targeted scenarios characterized by uniform devices and consistent data transmission sizes. In contrast, our study explores a broader spectrum, encompassing non-uniform devices, varying data transmission sizes, and fluctuating transmission rates to closely align with real-world scenarios.

\subsection{Graph Coloring}

Graph coloring, a key concept in graph theory, entails the assignment of colors to vertices (representing nodes or IoT devices) in a manner that guarantees adjacent vertices have unique colors, aiming to use the fewest colors (representing pilot sequences) possible. This approach holds notable advantages in addressing challenges posed by pilot contamination. The first benefit lies in its ability to optimize the economical use of limited pilot signal resources by minimizing color allocation.  Secondly, it maximizes the spatial separation between nodes or IoT devices, guided by predefined criteria such as interference, to ensure that adjacent nodes do not share the same color. 
\par In \cite{b5}, authors introduced a graph coloring-based technique to minimize pilot contamination within a cell-free M-MIMO system using a max k-cut problem. Max k-cut is NP-hard, making it computationally challenging to find optimal solutions, particularly for large graphs. In this paper, we formulate it as an optimization problem and use the ILP technique to find approximate near-optimal solutions. Acknowledging the computational intensity of the ILP, we introduce an efficient approach based on the binary search heuristic to find sub-optimal solutions within specified time constraints.

\begin{equation*}
I_{c,c'} = \sum\limits_{\substack{{c}=1}}^{C}\sum\limits_{\substack{{c'}=c}}^{C}\left| \frac{{\sum\limits_{\substack{\mathrm{k \in c_c}}}\mathrm{\beta}_{ik}^j}/{|c_c|}}{{\sum\limits_{\substack{\mathrm{k \in c_{c'}}}}\mathrm{\beta}_{jk}^j}/{|c_{c'}|}}\right| \tag {13}
\end{equation*}

\par We introduce a weight matrix to characterize interference between clusters from distinct cells, leveraging the large-scale fading coefficient $\beta$, which encompasses factors like geometric attenuation
and shadow fading. In the given equation, $I_{c,c'}$ represents the aggregate of average interference ratio originating from $c'_{th}$ cluster in the $i_{th}$ cell towards the $c_{th}$ cluster in the $j_{th}$ cell, and they could potentially share the same pilot signal. To account for the combined effect of multiple devices grouped in a cluster, we compute the average $\beta$ across the devices, where $|c_c|$ represents the $c_{th}$ cluster size.

\par In Algorithm 1, the input parameters consist of the number of cells (L), the number of clusters (C), and the interference matrix ($I_{c,c'}$), while the primary output is the computed threshold/interference value.  In lines 1 and 2, we establish the initial maximum and minimum values. Between lines 3 and 10, we calculate the number of devices included in the interference graph for a specified threshold/interference value. If this count falls below the C\_value, we proceed to invoke Algorithm 2 with that particular threshold value. In line 11, algorithm 2 returns the optimal pilot assignment and an exit flag corresponding to the given threshold value. Between lines 12 and 17, if the exit flag equals one, it signifies the attainment of an optimal solution within the allotted time frame (T). Subsequently, we increment the C\_value towards its maximum permissible value. Conversely, if the exit flag is not equal to one, indicating the absence of an optimal solution, we commence a decrementing process for C\_value toward its minimum permissible value.  In line 18, the algorithm concludes when the prev\_C\_value matches the current C\_value.

\begin{algorithm}
  \caption{Adaptive Threshold for ILP Using Binary Search Method}
  
  \KwIn{  $L,~ C,$~ $I_{c,c'}$,T }
  
  \KwOut{ Threshold Value}

  $max\_value = L \times C^{2}$\
  
  $[min\_value~ ,C\_value] = 1$\
  
  \For{j=1 to max\_value }{
    $\text{Threshhold} \gets 0.001$\
    
    \For{$i$ in range(1, 2000)}{
      ${count= \sum (I_{c,c'}>Threshold)}$\
      
      \If{$\text{count} < \text{C\_value}$}{
        \textbf{break}\
      }
      $\text{\textbf{else} Threshhold} \gets \text{Threshhold} + 0.001$\
    }
    ${Prev\_C\_value = C\_value}$\
    
    ${[{x, exitflag}] \gets \textbf{Algorithm 2} (C, I_{c,c'}, Threshold,T) }$
    
    \If{$\text{exitflag} == 1$}{
      $C\_value \gets  {(C\_value + max\_value)/2}$\
      
      $min\_value \gets Prev\_C\_value$\
    }
    \Else{
    $C\_value \gets  {(C\_value + min\_value)/2}$\
    
    $max\_value \gets Prev\_C\_value$\
    }
    
    \If{$Prev\_C\_value  == C\_value $}{ 
    \textbf{break}\
    }
  }
  
\end{algorithm}

  \subsection{Problem Statement}

This paper presents an optimization problem to minimize the required pilot overhead for a growing number of IoT devices within a cell while adhering to specified constraints. The problem is defined as follows, 

\begin{align*}
\tag{14a} & \mathbf{Minimize,~} \sum_{c=1}^{C}P_c \\
\tag{14b} & \mathbf{Subject ~to,} \sum_{c=1}^{C} C_{i,P_{c}} = 1 \quad \forall i=1,2,\ldots,L \\
&\tag{14c} \quad \quad  \quad \quad  ~~~  ~~C_{i,P_{c}} + C^{'}_{j,P_{c}} \leq 1 \quad \forall (i,j) \in  1,2,\ldots,L 
\end{align*}

In this formulation, the objective function aims to reduce the pilot overhead (number of orthogonal pilot signals) employed within a cell, where $P_c$ corresponds to the pilot sequence assigned to the $c_{th}$ cluster. The first constraint, expressed as $C_{i, P_{c}}=1$ where $i \in {1,2,\dots, L}$, relies on binary variables (0 or 1) to ensure that each cluster is dedicated exclusively to a single pilot sequence $P_c$. This binary variable signifies whether or not cluster $i$ is part of pilot sequence $P_c$. The second constraint ensures that if clusters $C$ and $C'$ are interconnected in the interference graph, where they could potentially share the same pilot sequence, they must be assigned different pilot sequences. This separation is vital to prevent interference between them.

\subsection{Integer Linear Programming (ILP)}
ILP is a mathematical optimization technique that seeks to optimize a linear objective function while adhering to a set of linear constraints, including integer constraints on decision variables. An inherent challenge in ILP is its NP-hard nature, which means that discovering an optimal solution can be computationally demanding and time-intensive. Furthermore, as problem complexity increases, the computational burden for solving ILP can grow exponentially. Therefore, this paper presents a binary search-based heuristic to address the ILP problem, aiming to attain a near-optimal solution within specified time constraints. 

\par In Algorithm 2 takes several inputs, including the number of clusters (C), an interference matrix $(I_{c,c'})$, a threshold value (Threshold), and a time constraint (T). In lines 1 and 2, the algorithm establishes an edge between the corresponding nodes when the interference matrix value exceeds the specified threshold. Between lines 3 and 5, the algorithm defines its objective function, aiming to minimize the number of pilot sequences allocated per cell. In lines 6 and 7, the algorithm formulates equality constraints. Between lines 8 and 13, the algorithm establishes the inequality constraints. In lines 15 and 16, the algorithm specifies the upper and lower bounds. Line 17 configures a time limit to halt the process if a solution is not found within a specified time T.  In line 18, the algorithm leverages a MATLAB ILP function. If it successfully finds an optimal solution x within a predefined time, it assigns the value '1' to the exit flag. Otherwise, it assigns a different value based on encountered errors before concluding the algorithm. This approach seeks to strike a balance between achieving optimality and managing computational time constraints.

\begin{algorithm}
  \caption{ Integer Linear Programming}
  \label{alg:graph_coloring_ip}
  \KwIn{ C, $I_{c,c'}, Threhsold,T$}
  
  \KwOut{ x, exitflag }

        \If{${I_{c,c'}} > \text{Threshold}$}{
          $\text{graph\_coloring}(c,c')=1$}

   $(n, m)=(L\times C,~size(graph\_coloring))$ \

  ${N \gets {n \cdot C + C}}$~~ \textbf{ Number of binary variables}\
  
  $f \gets [zeros(n \cdot C); ones(C)]$ ~~\textbf{Linear objective vector}\

  $A_{\text{eq}} \gets [ \text{repmat}( \text{eye}(n), 1, k) \text{zeros}(n, k)]$\
  
  $B_{\text{eq}} \gets \text{ones}(n)$ ~~\textbf{Equality constraints$: A_{\text{eq}},B_{\text{eq}}$}\
  
  \For{$i=1$ to $m$}{
    $A_{m}(i, \text{graph\_coloring}(i, :)) \gets 1$\
  }
  \For{$i=1$ to $C$}{
    $A_{y}(1+m \cdot (i-1) : i \cdot m, i) \gets 1$\
  }
  $A_{\text{ieq}} \gets [\text{kron}(\text{eye}(k), A_{m}) - A_{y}]$\
  
  $B_{\text{ieq}} \gets \text{zeros}(k \cdot m)$ ~\textbf{Inequality constraints$: A_{\text{ieq}},B_{\text{ieq}}$}\
  
  $intcon \gets 1:N$\
  
  $lb \gets \text{zeros}(N, 1)$ ~~\textbf{Lower bound}\
  
  $ub \gets lb + 1$~~\textbf{Upper bound}\
  
  $Options \gets{Stop~ time ~set ~to~ T ~seconds}$\
  
  $[x, \text{exitflag}] \gets \text{intlinprog}(f, intcon, A_{\text{ieq}}, B_{\text{ieq}}, A_{\text{eq}}, B_{\text{eq}}, lb, ub, options)$
  
  $\textbf{Return} \gets {x,~ exitflag}$\

\end{algorithm}

\section{Simulation}

We employed Matlab along with the communication toolbox, to simulate multi-cell M-MIMO systems in a 7-cell hexagonal configuration. We applied the wrap-around technique to counteract boundary effects. Within each cell, the maximum distance from a point to its base station is 125 meters. For a realistic wireless environment emulation, we considered 50 random channel realizations, with a path-loss exponent of \( \alpha = 3.76 \), shadow fading characterized by \( \sigma_{\text{sf}} = 10 \), a receiver noise level of -94 dBm, and an uplink transmit power of 20 dBm for devices. We assigned orthogonal pilot sequences to clusters within each cell and reused them in adjacent cells. We fixed a coherence block \( \tau_c \) at 200 samples.

\par Our investigation focused on the role of demand response in smart grid management. Demand response enhances the efficiency and utilization of smart grid resources, but this comes with a need for high-end communication infrastructure achievable through 5G-based IoT devices. To accurately model demand response management in our simulation, we considered a diverse range of devices varying in data sizes and transmission intervals. Devices were configured with data sizes of 500, 750, and 1000 bytes and had transmission intervals of 1, 2, or 3 seconds. All communications operated over a 12.5 KHz narrow bandwidth within the 902 MHz ISM spectrum.

\vspace{-0.4cm}
\begin{figure}[htp]
    \centering
    \includegraphics[width=8.40cm]{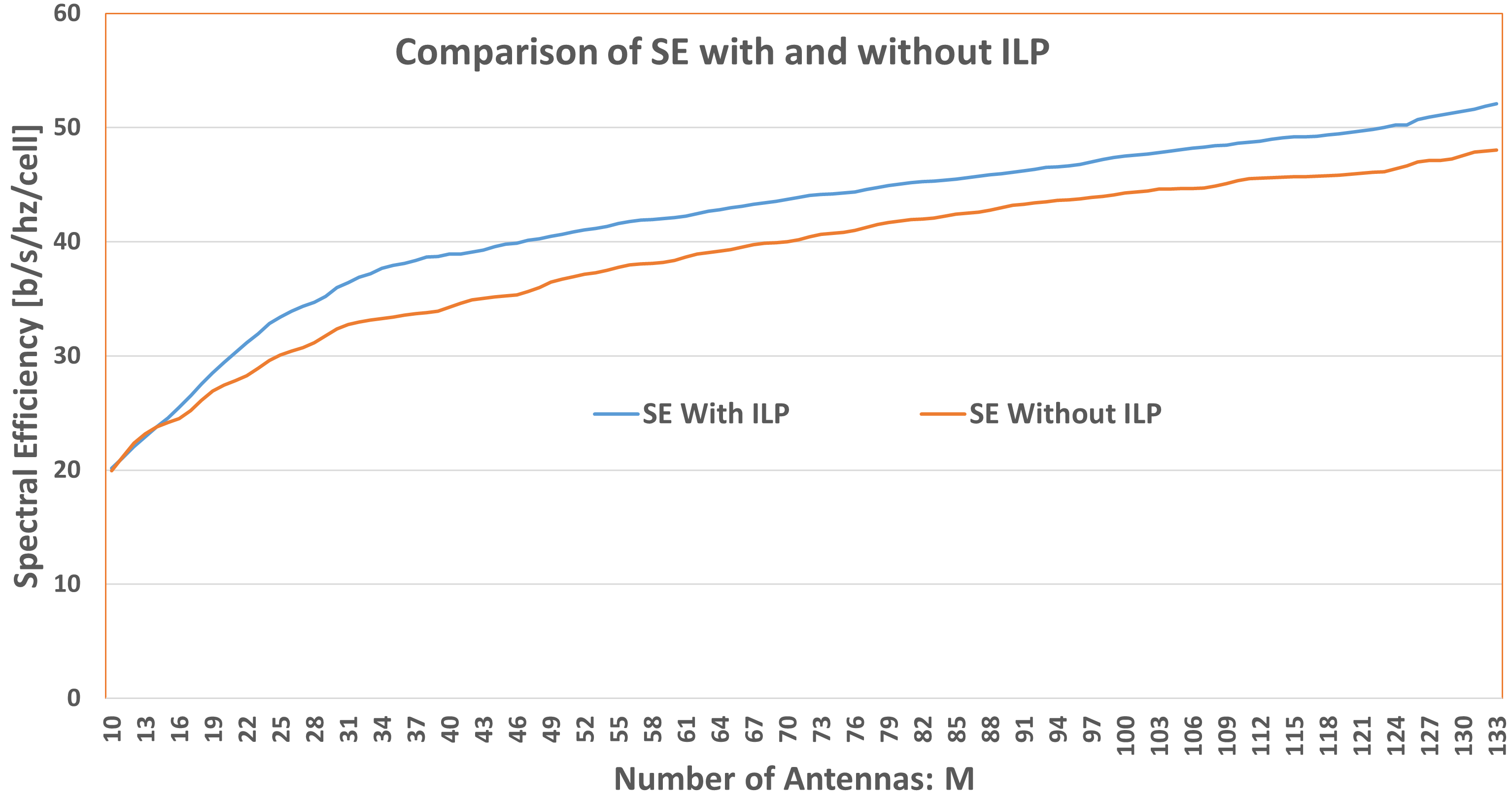}
   \vspace{-0.3cm}
    \caption{ Comparison of SE with and Without ILP}
    \label{fig:pilot}
\end{figure}
\vspace{-0.3cm}

\par Fig. 3 compares the SE with and without ILP. In this comparison, the number of antennae varies from 10 to 128 while keeping the number of IoT devices K fixed at 7. In the figure, both curves begin with an identical initial value of \(19.5 \, \text{b/s/Hz/cell}\). Notably, there is not a significant increase observed until the number of antennas reaches 20. This can be attributed to the fact that SE is directly proportional to the number of antennas. As the number of antennas increases, we witness a gradual increase in SE. Furthermore, we observe a steady rise, leading to the maximum difference of \(4.06 \, \text{b/s/Hz/cell}\) (about 14\%) between the two curves when the number of antennas is 40. The notable difference stems from ILP's effective pilot assignment, which reduces interference, thus enhancing SE. The difference remains consistent with minor fluctuations up to 128 antennas, where the SE for both with and without ILP measures at \(52.1 \, \text{b/s/Hz/cell}\) and \(48.04 \, \text{b/s/Hz/cell}\), respectively.

\vspace{-0.4cm}
\begin{figure}[htp]
    \centering
    \includegraphics[width=8.40cm]{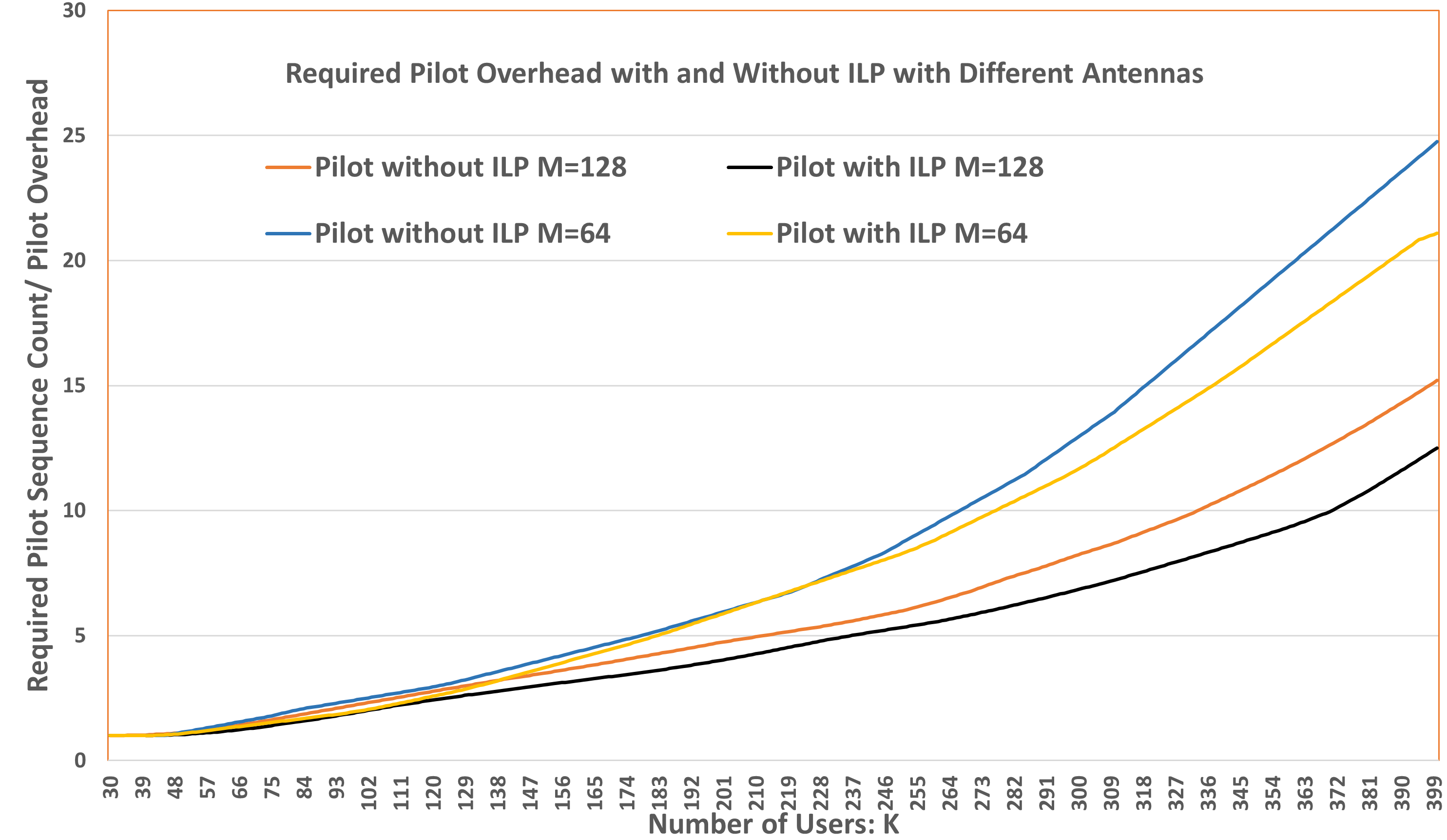}
   \vspace{-0.3cm}
    \caption{Required pilot overhead with and without ILP}
    \label{fig:pilot}
\end{figure}
\vspace{-0.4cm}

\par Fig. 4 compares the required pilot overhead (orthogonal pilot sequences count) per cell for different numbers of antennas. We sequentially select $K$ devices from C clusters, each with specific data sizes and transmission intervals, to optimize device scheduling. The aim is to optimize the number of required orthogonal pilot signals a cell would need to accommodate all selected devices while adhering to time constraints. From the figure, we can observe that until $K=120$, the required pilot count remains relatively stable, regardless of ILP usage or changes in antenna numbers. This stability is due to the small value of K, resulting in minimal impact on the overall required pilot count when enhancing SE using ILP. From $K=120$ onwards, a substantial difference in the required pilot sequence count becomes evident between the curves for antennas set at 64 and 128. With the antenna count fixed at 128, the required pilot overhead exhibits a clear gap between the ILP and non-ILP scenarios. As $K$ (the number of devices) escalates, the difference becomes more pronounced (about 18\%),  peaking at values of 12.5 for the with-ILP and 15.2 for the without-ILP scenarios when $K=400$. On the other hand, with the antennas set to 64, there is a more pronounced increase in the required pilot count comparatively when the antennas are at 128. The gap between the curves widens as $K$ grows, peaking at 21.08 for the with-ILP and 24.74 for the without-ILP scenarios when $K=400$. From Fig. 4 we notice a lower required pilot sequence for M = 128 compared to M = 64. This is because as the number of antennas increases, SE also increases, allowing for more users to be accommodated within the provided time slot, thus requiring fewer orthogonal pilot signals.

\section {Discussion and Future Research Directions}

This paper focuses on minimizing required pilot overhead per cell while effectively tackling pilot contamination and system scalability challenges in M-MIMO networks. The minimization of the required pilot sequence count is executed in two stages. First, using clustering, we assign a single orthogonal pilot sequence to each cluster. Then, we model the pilot assignment as a graph coloring problem, aiming to minimize the number of colors (or orthogonal pilot sequences) while ensuring unique sequences for connected nodes in the interference graph.

\par This also addresses the scalability of IoT devices in M-MIMO systems. As seen in Fig. 4, We can accommodate 400 devices using an average of only 12.5 pilot sequences per cell. This represents a notable advancement, especially when considering 200 samples per coherence block. The limited length of the coherence block previously restricted our ability to accommodate several devices as the pilot sequence length is always equal to or greater than the number of devices. Another significant contribution of this paper is mitigating pilot contamination through ILP. From the simulations, it is evident that ILP significantly enhances SE and offers substantial benefits in reducing the required pilot sequence count.

\par A potential future research avenue is to investigate the adaptability of the binary search algorithm, specifically by modifying its parameters in response to historical data trends. This adaptability might result in quicker processing and improved results. Moreover, expanding optimization techniques by merging ILP with methods like genetic algorithms or neural networks can enhance efficiency. This combination may optimize both computation speed and result accuracy. Future research should prioritize adaptive scheduling for diverse user dynamics, such as mobility in IoT-based M-MIMO systems, targeting enhanced resilience and efficiency.

\section{Conclusion}

5G sets the stage for enhanced performance and robust IoT interconnectivity. Massive MIMO is pivotal in 5G's advancement, boosting data speeds and refining network functions. This paper highlights the challenges and solutions in acquiring accurate CSI via pilot signals, underscoring the challenges from pilot signal reuse and resulting inter-cell pilot contamination. Our proposed ILP approach, coupled with a binary search-based heuristic, offers an innovative solution to these challenges, as evidenced by our simulation results. The groundbreaking capability to accommodate a large number of devices with a significantly reduced pilot overhead underscores the significance of our research. With the foundation laid, there is more work ahead. Future research should especially explore refining our algorithms' adaptability and integrating advanced optimization techniques, aiming to create a stronger and more streamlined IoT-based M-MIMO system.

\end{document}